\documentclass[preprint,
preprintnumbers,amsmath,amssymb,superscriptaddress]{revtex4-1}

\usepackage{fullpage,amsthm,amsmath,amsfonts,bm,times,color,bbm}
\usepackage{graphicx}
\usepackage[english]{babel}
\usepackage{mathtools,color}

\usepackage{graphicx}
\usepackage{float}
\usepackage{wrapfig}
\usepackage[figurename=Fig.,labelfont=bf]{caption}

\usepackage{subcaption}
\usepackage{hyperref}
\usepackage{romannum}
\usepackage{amsmath}
\usepackage{amssymb}
\usepackage{multirow}
\usepackage{booktabs}
\usepackage{makecell}
\usepackage[flushleft]{threeparttable}

\usepackage{soul} 

\newcommand{\chev}{$\mathrm{CV}$}

\newcommand{\el}{El Ni\~{n}o}
\newcommand{\la}{La Ni\~{n}a}

\makeatother

\makeatletter

\begin{document}


\pagenumbering{arabic}

\title{Emergence of universal scaling in weather extreme events}%

\author{Qing Yao}%
\affiliation{School of Systems Science/Institute of Nonequilibrium Systems, Beijing Normal University, 100875 Beijing, China}
\affiliation{Blackett Laboratory and Centre for Complexity Science, Imperial College London, South Kensington Campus, London SW7 2AZ, United Kingdom}

\author{Jingfang Fan}%
\email[Corresponding authors:]{jingfang@bnu.edu.cn}
\affiliation{School of Systems Science/Institute of Nonequilibrium Systems, Beijing Normal University, 100875 Beijing, China}
\affiliation{Potsdam Institute for Climate Impact Research, 14412 Potsdam, Germany}

\author{Jun Meng}%
\affiliation{School of Science, Beijing University of Posts and Telecommunications, 100876 Beijing, China}
\affiliation{Potsdam Institute for Climate Impact Research, 14412 Potsdam, Germany}

\author{Valerio Lucarini}
\affiliation{Department of Mathematics and Statistics, University of Reading, Reading, RG6 6AX, United Kingdom}
\affiliation{Centre for the Mathematics of Planet Earth, University of Reading, Reading, RG6 6AX, United Kingdom}
\affiliation{School of Systems Science/Institute of Nonequilibrium Systems, Beijing Normal University, 100875 Beijing, China}

\author{Henrik Jeldtoft Jensen}%
\affiliation{Department of Mathematics and Centre for Complexity Science, Imperial College London, South Kensington Campus, London SW7 2AZ, United Kingdom}
\affiliation{ Institute of Innovative Research, Tokyo Institute of Technology, 4259, Nagatsuta-cho, Yokohama, 226-8502, Japan}

\author{Kim Christensen}%
\affiliation{Blackett Laboratory and Centre for Complexity Science, Imperial College London, South Kensington Campus, London SW7 2AZ, United Kingdom}

\author{Xiaosong Chen}%
\affiliation{School of Systems Science/Institute of Nonequilibrium Systems, Beijing Normal University, 100875 Beijing, China}

\begin{abstract}
    The frequency and magnitude of weather extreme events have increased significantly during the past few years in response to anthropogenic climate change. However, global statistical characteristics and underlying physical mechanisms are still not fully understood. Here, we adopt a statistical physics and probability  theory based method to investigate the nature of extreme weather events, particularly the statistics of the day-to-day air temperature differences. These statistical measurements reveal that the distributions of the magnitudes of the extreme events satisfy a universal \textit{Gumbel} distribution, while the waiting time of those extreme events is governed by a universal \textit{Gamma} function. Further finite-size effects analysis indicates robust scaling behaviours. We additionally unveil that the cumulative distribution of logarithmic waiting times between the record events follows an \textit{Exponential} distribution and that the evolution of this climate system is directional where the underlying dynamics are related to a decelerating release of tension. The universal scaling laws are remarkably stable and unaffected by global warming. Counterintuitively, unlike as expected for record dynamics, we find that the number of quakes of the extreme temperature variability does not decay as one over time but with deviations relevant to large-scale climate extreme events.   
    Our theoretical framework provides a fresh perspective on the linkage of universality, scaling, and climate systems. The findings throw light on the nature of the weather variabilities and could guide us to better forecast extreme events.
\end{abstract}
    \date{\today}

\maketitle
\section*{Introduction}
Natural or social systems far from equilibrium often show complex dynamic behaviours. Nevertheless, increasing empirical evidence supports that the fields as diverse as physics~\cite{odor2004universality,fan2020universal}, biology~\cite{munoz2018colloquium,di2018landau}, ecology~\cite{sole2007scaling,ross2021universal}, human mobility~\cite{song2010modelling}, economics, and financial markets~\cite{bouchaud1990anomalous,mantegna1995scaling} may all, at least up to a certain degree of approximation, follow the principles of scale invariance and universality~\cite{stanley2000scale}. Scaling functions characterise the critical phenomena of phase transitions in physical systems. The following mounting evidence demonstrates these functions are also powerful tools for understanding and predicting the behaviours of various complex systems~\cite{sornette2006critical,west2017scale}, such as the climate system~\cite{ashkenazy2003nonlinearity,fan2021statistical} and earthquakes~\cite{bak2002unified,saichev2006universal,corral2004long,livina_memory_2005}.

Understanding the nature and the statistical properties of climate variability across multiple spatial and temporal scales and its relationship with the ongoing climate change is a key goal of climate science \cite{ghil2020physics,franzke2020structure}. Evaluating how climate change will impact the tail of the distribution of climatic variables is crucial  to evaluate the impacts of climate change, as extreme events are often conducive to risks to human and environmental welfare~\cite{katz1992extreme,karl1995trends,Easterling2000,Ipcc2012}. A great deal of attention has been devoted to understanding to what extent it is possible to attribute individual extreme events to man-made climate change ~\cite{Allen2003,trenberth2015attribution,Otto2017}.  The World Meteorological Organization (WMO) reported that weather-related disasters have increased over the past 50 years, causing more damage, but early warnings save lives~\cite{WMO}. According to WMO, economic losses have increased sevenfold from the 1970s to the 2010s, but, thanks to the improved early warnings and disaster management, the number of deaths decreased almost three-fold over the 50-year period~\cite{WMO}.
In particular, a growing body of literature has identified the social impacts of large temperature variability at different time scales in areas like human health~\cite{Poumadere2005,shi2015impacts}, crop production~\cite{wheeler2000temperature}, and economic growth~\cite{kotz2021day}. Conversely, extremes reveal essential aspects of the dynamics of climate because there is a fundamental connection between suitably defined extreme events and dynamical processes of the system generating them \cite{albeverio2006extreme,lucarini2016extremes,Galfi2017,Faranda2017,Hochman2019,Galfi2021}. Additionally, thanks to the presence of couplings acting on different temporal and spatial scales, the climate system, behaving similarly to a complex network, can feature the spread of the effects of extreme events, leading to cascades of risks~\cite{helbing2013globally}. 
This study aims at advancing our understanding of extreme events in the climate system using the viewpoint provided by 
statistical physics and probability theory. Specifically, we will take advantage of extreme value theory \cite{Leadbetter1983,Katz2002,Ghil2011,coles2001introduction,lucarini2016extremes}, which provides a very powerful and flexible framework for studying the tails of the distribution of random variables. We will focus on near-surface temperature fields and consider different sites covering the globe. Instead of looking, as often done, at daily temperature records, we will study the extreme events of the day-to-day temperature change (both positive and negative, as well as their absolute value). We remark that extreme positive and negative events are, in general, associated with different physical processes: cold and warm weather fronts are fundamentally different at the dynamical level~\cite{Holton2004}. Looking at this climatic variable, which is the discrete approximation of the time derivative of the temperature field, amounts to studying one component of the local energy budget of near-surface air masses (the missing component being associated with moisture content). We remark that a recent study  looked at the spatial statistics of daily temperature variability measured as the intramonthly standard deviation of daily surface temperature. The authors found a relevant impact of CO2 increase on this type of measure of weather variability~\cite{Kotz2021b}. As discussed below, our study uses a different measure of daily temperature variability, which seems better suited to representing the challenges of rapid adaptation of human and environmental systems to rapid temperature variations.


{\subsection*{Extreme Events}}
We study the daily temperature variations using the differences in temperature between two adjacent discrete time periods, that is $\mathrm{CV}_{i,d} = x_{i,d} - x_{i,d-1} , ~ i = 1, ~\dots, N, d = 2~\dots D$, where $x_{i,d}$ is air temperature recorded for day $d$ at site $i$; $D$ is the total number of days of the recorded data, and the analysis spans 43 years, from 1979 to 2021. $N$ is the total number of sites ($73 \times 144$) covering the entire Earth based on the ECMWF Reanalysis 5th Generation (ERA5)~\cite{hersbach2020era5} dataset, see Data availability for more details. Furthermore, the absolute values of the changes $\lvert \mathrm{CV}_{i,d} \rvert = \lvert x_{i,d} - x_{i,d-1} \rvert$ provides a measure of the strength of weather variability~\cite{karl1995trends}.

\begin{figure}[!h]
    \centering
    \includegraphics[width=1\textwidth]{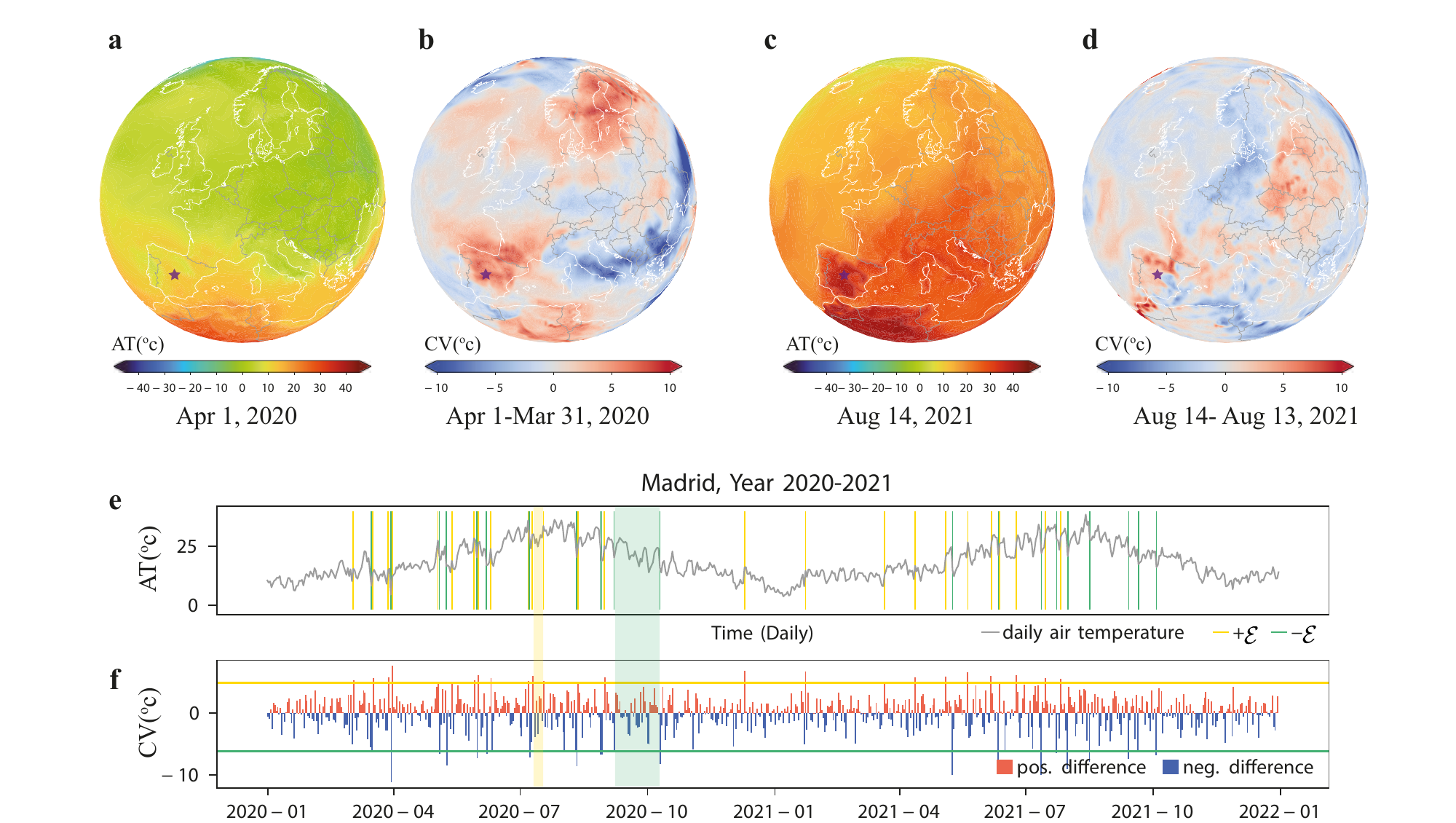}
    \caption{\textbf{The illustration of weather extreme events defined for day-to-day air temperature variability. a} and \textbf{c} are air temperatures of single days (April 1, 2020 and August 14, 2021) visualised through nearside perspective projection; \textbf{b} and \textbf{d} are the air temperature differences between the next day and the chosen days. We choose a single site in Madrid ($40.25^{\circ}$N,  $3.75^{\circ}$W), Spain, to illustrate the definitions of extreme events for day-to-day air temperature variability. 
     \textbf{e} are the daily air temperatures (AT) from January 2020 to Dec 2021. \textbf{f} is the day-to-day temperature differences, $\mathrm{CV}$, and the red bars and navy bars represent the positive and negative differences between the two adjacent days, respectively. The horizontal yellow (sea green) lines are the threshold of $97.5\%$ positive (negative) changes. The day-to-day temperature differences are greater (lower) the thresholds are marked by vertical yellow (sea green) lines, which indicate the happening of the extreme events defined for the given thresholds. The width of the yellow (sea green) rectangular is the waiting time between positive (negative) extremes. }\label{fig:def_extreme}
\end{figure}

A clear illustration of these definitions for one selected site is given in Fig.~\ref{fig:def_extreme}, which shows a 2-year time slice of daily temperature reading and day-to-day temperature variations recorded in Madrid ($40.25^{\circ}$N,  $3.75^{\circ}$W), Spain. The plot shows that extreme day-to-day temperature differences are more common before the summertime, followed by high-temperature days from June to August (the summertime of the Northern Hemisphere and the sites in the South Hemisphere follows similar seasonal trends, see Fig.~S1 in SI).
This is related to the well-known stronger weather variability in winter versus summer, associated with the presence of  stronger baroclinic conversion processes~\cite{Holton2004}.

To investigate the approach to the limiting behaviour in the statistics of temperature differences defined by extreme value theory, we apply the Peaks Over Threshold methods~\cite{coles2001introduction} to the normalised temperature positive and negative differences for all sites separately. We show the dependence of the estimates of the shape parameter and of the modified scale parameter on the threshold value for sites on land and ocean in Fig.~S2 in SI. The distribution of the estimates of shape parameters is broader over the oceans, which means the asymptotic behaviours are less accurately established for ocean sites. However, most of the shape parameters of oceans and lands fall within a reasonably small range of $[-0.1, 0]$, and the asymptotic regime is established for quantiles as low as $(60\%, 67.5\%)$ because the estimates of the scale parameters and of the modified scale parameter become stable as we consider higher quantiles \cite{coles2001introduction,galfi2017convergence}. This regime is considerably lower than what is usually found when looking at temperature extremes, see, e.g.~\cite{Zahid2017}, which is encouraging in terms of applicability of extreme value theory for day-to-day temperature variability also in conditions where data availability is limited.

We consider all sites on the entire Earth together. With this consideration, we have lost the structure in space and in the magnitude scale; nonetheless, the change in the process properties with $q$ within the extreme regime will allow us to recover some information. A positive extreme event set $\mathcal{E}^{+}_{i}(q)$ is defined when the value~\chev~exceeds a chosen threshold associated with a quantile $q$. Then, we define the magnitude of the positive extreme temperature difference as the normalised value of the event, that is $m^{+}_{ev} = \frac{\mathrm{CV}_{ev^{+}}}{\sigma_{CV_{ev^{+}}}},~ ev \in  \mathcal{E}^{+}_{i}$,
where $\sigma_{CV_{ev^{+}}}$ is the standard deviation of the extreme part, $\mathrm{CV}_{ev^{+}}$ for site $i$. Similarly, a negative extreme event set $\mathcal{E}^{-}_{i}(q)$ is defined when the value~\chev~is smaller than a chosen quantile $q$. Then, for simplicity, we define the magnitude of the negative extreme temperature difference as the minus normalised value of the event, that is $m^{-}_{ev} = - \frac{\mathrm{CV}_{ev^{-}}}{\sigma_{CV_{ev^{-}}}},~ ev \in  \mathcal{E}^{-}_{i}$. After taking the minus of the values, the $m^{-}_{ev}$ will be positive, which is comparable to the $m^{+}_{ev}$. Likewise, we define for the absolute values, $m^{abs}_{ev} =  \frac{\lvert  \mathrm{CV} \rvert _{ev^{abs}}}{\sigma_{CV_{ev}^{abs}}},~ ev \in  \mathcal{E}^{abs}_{i}$.

The waiting time $\tau_{i} \in \mathrm{T}_{i}$ of the extreme events in $\mathcal{E}_{i}$ is the number of days between two consecutive events~\cite{boers2019complex}. Therefore, we can define four types of waiting times:
$\tau^{+}$ ($\tau^{-}$, $\tau^{abs}$) is the waiting time between positive (negative, absolute) extreme temperature difference, \textit{i.e.} the waiting time in $\mathcal{E}^{+}$ ($\mathcal{E}^{-}$, $\mathcal{E}^{abs}$). The fourth type $\tau^{+-}$ ($\tau^{-+}$ interchangeable) is the number of days between extreme positive and extreme negative change events (or extreme negative and extreme positive), \textit{i.e.} the waiting time in $( \mathcal{E}^{+} \cup \mathcal{E}^{-})$ and the detailed description can be found in Fig.~S5 in SI.

\section*{Results}

\subsection*{Universal scaling for magnitudes of extreme temperature differences}


Here, we focus on the distributions of the dimensionless magnitudes $m^{+}_{ev}$ and $m^{-}_{ev}$ for all sites on the Earth, that is, the positive and negative extreme parts of $\mathrm{CV}$. We choose $q = 90\%, 92.5\%, 95\% $ and $97\%$ as threshold quantiles within the extreme regime. 

\begin{figure}[!h]
    \centering
    \includegraphics[width=1\linewidth]{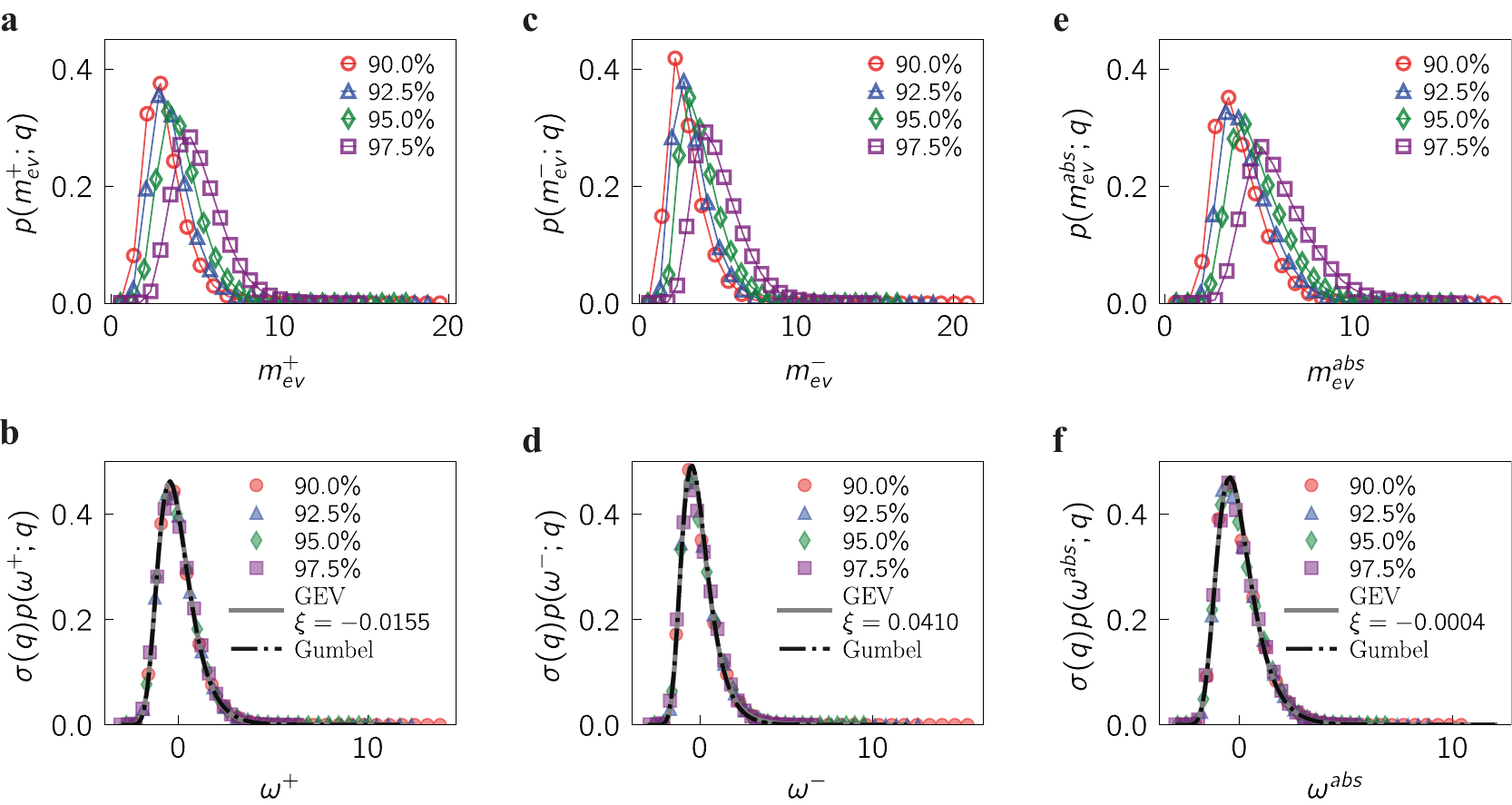}
    \caption{The magnitudes of extreme day-to-day temperature, \textbf{a}, \textbf{c} and \textbf{e} are probability density distribution for different chosen quantiles (thresholds) for positive, negative and absolute extreme events respectively; \textbf{b}, \textbf{d} and \textbf{f} are universal scaling functions. The fitting method is Maximum Likelihood Estimation (MLE), and we use the Kolmogorov–Smirnov(KS) test for the cumulative distribution of $95\%$. The fitting parameters are shown in Tab.~\ref{tab:parameters}. The fitted lines by GEV are plotted in grey, and the lines which presented Gumbel distribution are in dashed black lines. These two distributions are hardly distinguishable in \textbf{b}, \textbf{d} and \textbf{f} for PDF. The associated cumulative distribution functions are in SI, see Fig.~S4.}
    \label{fig:atch_m}
\end{figure}

The normalised PDF for $m^{+}_{ev}$ and $m^{-}_{ev}$, $p(m^{+}_{ev};q)$ and $p(m^{-}_{ev};q)$ are shown in Figs.~\ref{fig:atch_m}a and c, respectively. The PDF satisfies a single peak distribution and depends on our chosen threshold quantiles $q$.  The distributions of $m^{+}_{ev}$ and $m^{-}_{ev}$ are expected to take the following scaling forms, 
\begin{equation} \label{Scaling:1}
    p(m^{+}_{ev};q) = \frac{1}{\sigma(q)}f^{+} \left(\frac{m^{+}_{ev} - \langle m^{+}_{ev} \rangle}{\sigma(q)} \right),
\end{equation}
and 
\begin{equation} \label{Scaling:2}
    p(m^{-}_{ev};q) = \frac{1}{\sigma(q)}f^{-} \left( \frac{m^{-}_{ev} - \langle m^{-}_{ev} \rangle}{\sigma(q)} \right),
\end{equation}
where $f^{+}$ and $f^{-}$ are two universal scaling functions, $\langle m^{+}_{ev} \rangle$ ($\langle m^{-}_{ev} \rangle$) and $\sigma(q)$ is the mean value and standard deviation associated with quantile $q$. We then define two rescaled parameters
$\omega^{+} = \frac{m^{+}_{ev} - \langle m^{+}_{ev} \rangle} {\sigma(q)}$ and $\omega^{-} = \frac{m^{-}_{ev} - \langle m^{-}_{ev} \rangle} {\sigma(q)}$.
Combining the proposed scaling forms, Eqs.~\eqref{Scaling:1} and \eqref{Scaling:2}, with computing values $\sigma(q)$ of different $q$, yields
collapse onto the respective universal scaling functions, as shown in Figs.~\ref{fig:atch_m}b and d. In particular, we find that both  $f^{+}$ and $f^{-}$ are well described by the \textit{Gumbel} extreme function, 
\begin{equation} \label{eq:3}
    f^{+}(z^{+}) = \frac{1}{\beta}e^{-e^{-z^{+}}-z^{+}},
\end{equation}
and 
\begin{equation} \label{eq:4}
    f^{-}(z^{-}) = \frac{1}{\beta}e^{-e^{-z^{-}}-z^{}},
\end{equation}
where $z^{+} = (\omega^{+}-\mu^{+})/\beta^{+}$ and $z^{-} = (\omega^{-}-\mu^{-})/\beta^{-}$. $\mu^{+}$ ($\mu^{-}$) is the location parameter (the mode) and $\beta^{+}$ ($\beta^{-}$) is the scale parameter.

The physical mechanism underlying our above conjecture is based on the Generalised Extreme Value (GEV) theory, see Eq. \eqref{eq:1}.
When we fit our rescaled collapsed data with the GEV distributions (see Methods for the fitting process), we find that the shape parameter $\xi$ is very close to zero (see Tab.~\ref{tab:parameters}). The GEV functions with a small $\xi$ (Eq. \eqref{eq:1}) and Gumbel function (Eqs. \eqref{eq:3} and \eqref{eq:4}) are 
complemented in Figs.~\ref{fig:atch_m}b, d and f. They are hardly distinguishable, which indicates that the $f^{+}$ ($f^{-}$) can be well approximated by a Gumbel distribution. We analyse the cumulative distribution function (CDF), and the data are also collapsed together. The results are consistent with the PDF's, see Fig.~S4 in SI.

\subsection*{Universal scaling for waiting time between extreme events}

\begin{figure}[!h]
    \centering
    \includegraphics[width=1\linewidth]{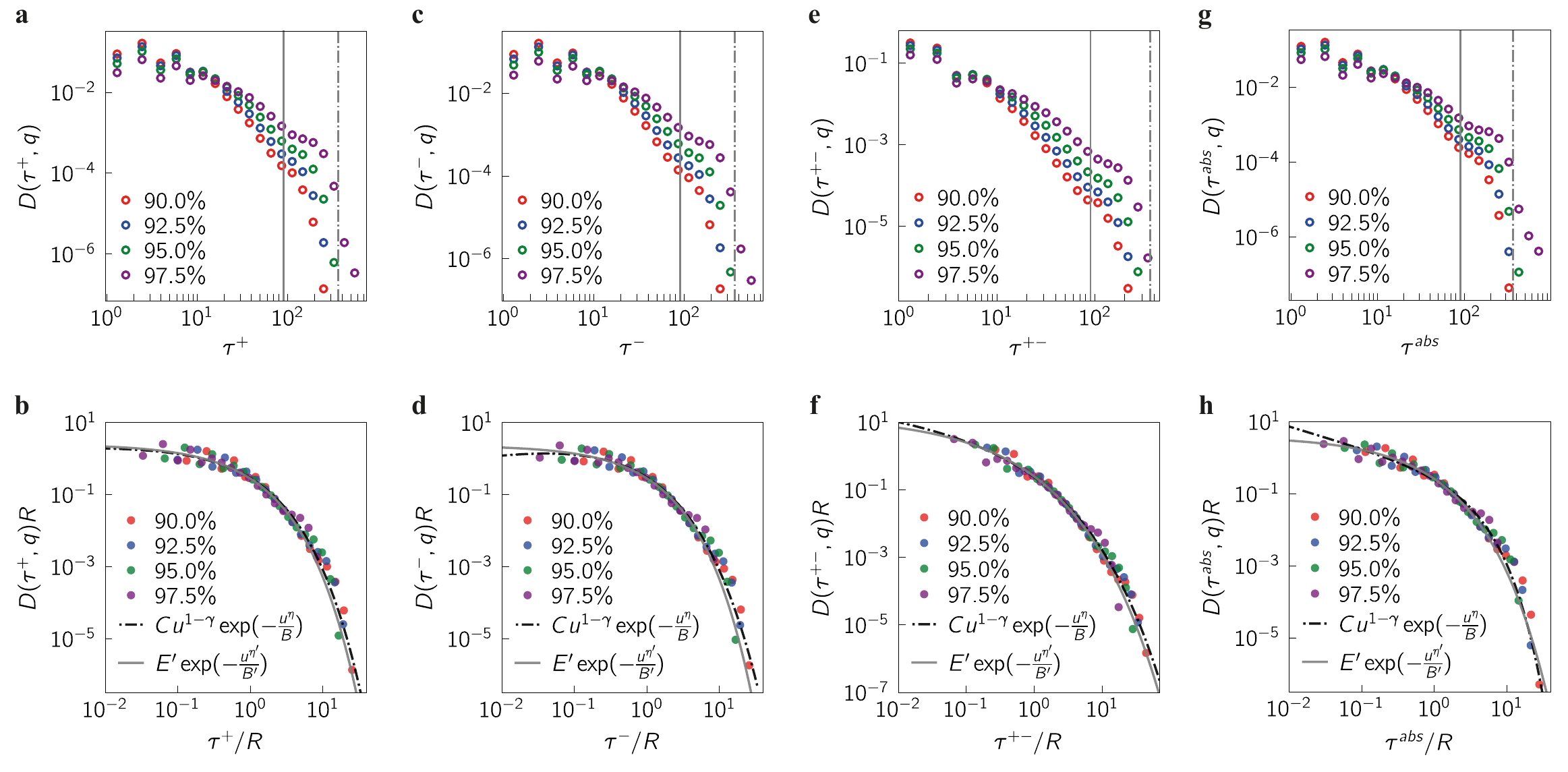}
    \caption{\textbf{Waiting time of extreme day-to-day temperature differences.} The four columns are $\tau^{+}$, $\tau^{-}$, $\tau^{+-}$ and $\tau^{abs}$, respectively. The fitting parameters are in Tab.~\ref{tab:parameters}. \textbf{a}, \textbf{c}, \textbf{e} and \textbf{g} are log-binned probability density distribution for different chosen quantiles with the base of $1.3$~\cite[Ch.\ 7, p. 233]{pruessner2012self}. The dashed grey vertical lines indicate $365$ days, while the solid grey vertical lines mark $90$ days. \textbf{b}, \textbf{d}, \textbf{f} and \textbf{h} are the universal scaling functions. The fitting parameters for the generalised Gamma function, Eq.~(\ref{Scaling:8}), are summarised in Tab.~\ref{tab:parameters}. The statistics of the mean value $R$ of waiting times and fitting parameters for exponential function can be found in Tab.~S1 in SI. }\label{fig:atch_WT}
\end{figure}

The waiting time for extreme events is also crucial for understanding the risk management of systems. As we mentioned before, for a given site and quantile $q$, we introduced four types of waiting time $\tau^{+}$, $\tau^{-}$, $\tau^{+-}$ ($\tau^{-+}$) and $\tau^{abs}$. We first consider their PDFs: $D(\tau^{+}; q)$, $D(\tau^{-}; q)$, $D(\tau^{+-}; q)$ and $D(\tau^{abs}; q)$. Since the multiple time scales are involved from days to many years, we work with the logarithm of $\tau$. Figures~\ref{fig:atch_WT}a, c, e and g show the results for different $q$ of the waiting time $\tau^{+}$, $\tau^{-}$, $\tau^{+-}$ and $\tau^{abs}$, respectively. We define $R$ as the mean value of $\tau$ for a certain threshold $q$ and rescale distributions of different thresholds with the mean rate $1/R$. Figures~\ref{fig:atch_WT}b, d, f and h show the rescaling results, and this data collapse implies that (take $\tau^{+}$ as an example)
\begin{equation}\label{Scaling:7}
    D(\tau^{+}; q) = \frac{1}{R} \mathcal{G} \left( \tau^{+}/R \right),
\end{equation}
where $\mathcal{G}$ is another universal scaling function. Similarly, we can also obtain the universal scaling functions for $\tau^{-}$, $\tau^{abs}$ and $\tau^{+-}$.  Motived by Ref. \cite{corral2004long} for earthquakes, we find that the scaling function $\mathcal{G}$ can be well fitted by a generalised \textit{Gamma} distribution,
\begin{equation}\label{Scaling:8}
    \mathcal{G}(u;\gamma,\eta,B,C) = C\frac{1}{u^{1-\gamma}}\exp \left(-\frac{u^{\eta}}{B}\right ),
\end{equation}
where $\gamma, \eta, B$ and $C$ are fitting parameters. All fitting values are summarised in Tab.~\ref{tab:parameters}.

The high-quality of the fit obtained using the generalised Gamma distribution is, at first sight, surprising because the fitting parameters indicate relevant deviations from the exponential behaviour, which is instead expected in the case of chaotic flows, under the hypothesis of stationarity \cite{lucarini2016extremes}. The weather is indeed chaotic and one expects no long time memory. But, on the other hand, the climatic data we consider are not stationary because one faces - on the time scales of relevance for the waiting times - the strong modulations due to the daily and seasonal cycle. Our data analysis procedure automatically filters out the former because we consider daily averaged data \footnote{Note that in continental and arid areas, day-to-night excursions of more than 30 $^\circ C$ are far from atypical}. As for the latter, we observe that if we restrict the analysis to waiting times shorter than approximately 90 days, thus within the sub-seasonal range (these contribute to approximate or over $90\%$ of the measurements), the statistics of waiting times is exponential to a high degree of accuracy. The exponential law is indicated in grey in Figs. \ref{fig:atch_WT} (b, d, f, h) and the statistics are given in Tab.~S1 in SI.

\begin{table}[]

            \begin{threeparttable}
                \centering
                \begin{tabular}{ccccccc}
                      &             \multicolumn{5}{c}{\bf Magnitude}  \\
                      \hline\hline
                      &         \boldmath $\xi$                                & \boldmath $\mu$                               & \boldmath  $\beta$                             & \boldmath $R^{2}$& \bf KS-Dn(\boldmath $95\%$)  \\\hline

                      \multicolumn{1}{c|}{\boldmath $m^{+}_{ev}$}& $-0.0155$ $(0.0001)$ & $-0.4487$ $(0.0001)$ & $0.7962$ $(0.0001)$ &  $0.997$ & $0.56$***  \\\hline

                      \multicolumn{1}{c|}{\boldmath $m^{-}_{ev}$}
                    & $0.0410$ $(0.0001)$  & $-0.4647$ $(0.0001)$ & $0.7472$ $(0.0001)$ &  $0.992$ & $0.65$***   \\\hline

\multicolumn{1}{c|}{\boldmath $m^{abs}_{ev}$}
                    & $-0.0004$ $(0.0001)$  & $-0.4538$ $(0.0001)$ & $0.7828$ $(0.0001)$ &  $0.994$ & $0.64$***   \\\hline

                      &         &       &       &   &  \\

                      &            \multicolumn{5}{c}{\bf Waiting time}                  \\
                      \hline\hline
                      &    \boldmath $\gamma$                           & \boldmath $\eta$                                 & \boldmath $B$                                 & \boldmath $C'$   & \boldmath $R^{2}$                                                        \\
                      \hline
                      \multicolumn{1}{c|}{\boldmath $\tau^{+}$}  &
                       \centering$1.04$ $(0.36)$ & \centering $0.56$ $(0.11)$       & \centering$0.44$ $(0.23)$      & \centering$1.03$ $(1.24)$   &    $0.966$             \\\hline

                      \multicolumn{1}{c|}{\boldmath $\tau^{-}$}  & \centering$1.36$ $(0.50)$ & \centering$0.47$ $(0.11)$       & \centering$0.29$ $(0.17)$      & \centering$2.22$ $(2.03)$    &                $0.964$               \\\hline

                      \multicolumn{1}{c|}{\boldmath $\tau^{+-}$} & \centering$0.73$ $(0.54)$ & \centering$0.39$ $(0.11)$       & \centering$0.33$ $(0.25)$      & \centering$1.59$ $(2.37)$    &     $0.982$                         \\\hline
                       \multicolumn{1}{c|}{\boldmath $\tau^{abs}$} & \centering$0.40$ $(0.18)$ & \centering$0.85$ $(0.13)$       & \centering$1.49$ $(0.72)$      & \centering$-0.75$ $(0.41)$    &     $0.964$                         \\
                      \hline
                \end{tabular}
                \end{threeparttable}
                        \caption{\textbf{Table of fitting parameters.} The numbers in the brackets are the standard errors of the nonlinear fitting methods. We fit the magnitudes with the Maximum Likelihood Estimator (MLE) and then test the goodness of fit using the $95\%$ data through the Kolmogorov–Smirnov (KS) test. `***'means $p<0.001$. The waiting time is fitted in the log scale through the least square errors method; therefore, the $C = \exp C'$. The detailed information on the fitting and test are in the Methods.}\label{tab:parameters}
            \end{table}

\subsection*{Universal scaling of finite-size analysis}

To better understand whether the scaling behaviour change with the system size and regions, we investigate the finite-size effects of the universal scaling. Our previous analysis of the dataset is based on the entire Earth with a resolution of $2.5^{ \circ}$, yielding $N = 73\times144 = 10512$ nodes. Note that the nodes are not approximately homogeneously covering the entire globe due to the Earth’s ellipsoid shape. To remove the non-homogeneity, we perform the procedure developed in Ref.~\cite{meng_percolation_2017}. We define the resolution (in degree latitude) at the Equator as $r_{0}$ and calculate the number of nodes $n_{0}=360 / r_{0}$. Then the number of nodes in latitude $mr_{0}$ is $n_{m}$ $=n_{0} \cos \left(mr_{0}\right)$, where $m \in\left[-90 / r_{0}, 90 / r_{0}\right]$. The total number of nodes is $N'=\sum_{m=0}^{m=90 / r_{0}} 2 n_{m}-n_{0}$. Here, we choose $r_{0}$ to be $2.5^{ \circ}$, which yields $N'=6570$. Furthermore, we divide the globe (6570 nodes) into three `Tropics' ($25^{ \circ}$S to $25^{ \circ}$N), `Mid-latitude' ($60^{ \circ}$S to $27.5^{ \circ}$S and $27.5^{ \circ}$N to $60^{ \circ}$S), as well as `High latitude' regions ($90^{ \circ}$S to $62.5^{ \circ}$S and $62.5^{ \circ}$N to $90^{ \circ}$N), see Fig.~S6 in SI.

The same analysis is performed on the above three regions, and identical results hold, as Fig.~S7 and S8 illustrate the universal scaling of magnitudes and waiting times, respectively. We find that the universal scaling functions for magnitudes are robust and very similar to the results of the entire Earth (Fig.~\ref{fig:atch_m}); The universal scaling phenomenon of the magnitudes and waiting time are very well described by the Gumbel and the Gamma functions, but with different fitting parameters for the three regions.

\subsection*{Evolution of the climate system}
The weather/climate system is not an equilibrium system, and \textit{`Quakes'} is a feature that can characterise complex system dynamics ~\cite{sibani1993slow,jensen2013stochastic}, which has been applied in other non-equilibrium systems, including physical systems and ecology systems~\cite{anderson2004evolution}. The quakes of complex systems gradually change both the physical and statistical properties of the system. The \textit{record} is defined as the largest value of a time signal obtained up to time $t$. The quake times $t_{k}$ is the time where the record reaches a higher value, indicating when the record jumps. It is shown that the quake event times of system variable $t_{k}$ undergoes record statistics satisfy the \textit{log-Poission} distribution~\cite{anderson2004evolution}.
In particular, the logarithmic waiting time between quakes defined as $\tau_{k} = \log(t_{k}) - \log(t_{k-1})$, follows a common cumulative distribution function:
\begin{equation} \label{EQ:CCD}
    P(\tau_{k}) = 1 -  e^{-\tau_{k}}.
\end{equation}

\begin{figure}[!h]
    \centering
    \includegraphics[width=0.9\linewidth]{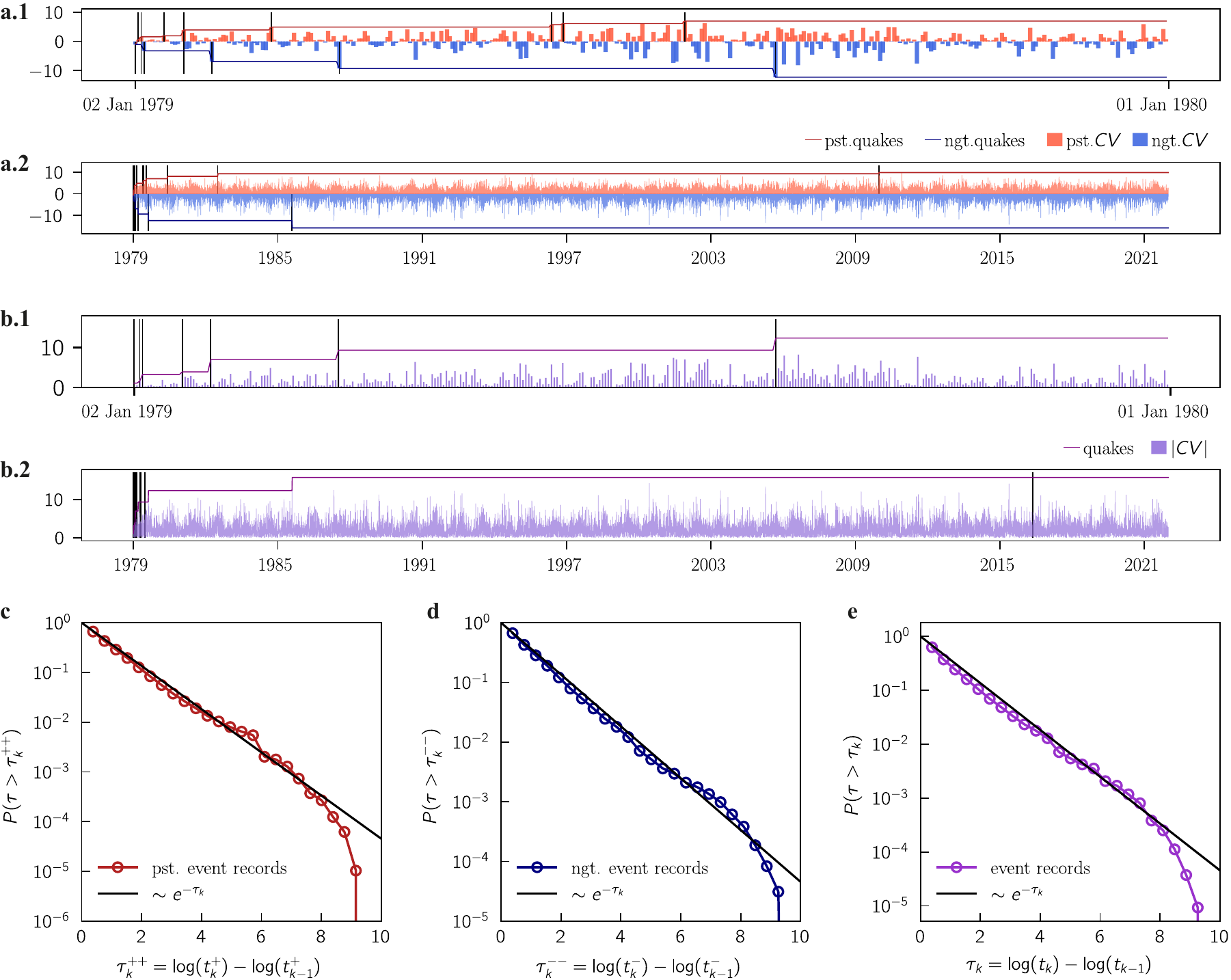}

    \caption{\textbf{The quake times of the temperature differences and CDF of their logarithmic waiting time for a site around Madrid, which is from the same time signal as Fig.~\ref{fig:def_extreme}e.} \textbf{a}, The red bars represent the positive changes, $\mathrm{CV}>0$, while the blue bars represent negative changes, $\mathrm{CV}<0$. The red line represents the largest positive value of $\mathrm{CV}>0$ up to time $t$, and the blue line represents the largest negative value (i.e. the smallest values) of $\mathrm{CV}<0$ up to time $t$. The upper plot is the zoom-in part of the year 1979. \textbf{b}, The purple bars represent the absolute values of \textbf{a}, the purple line indicates the largest value of $\rvert\mathrm{CV}\rvert$ up to time $t$, and the \textbf{b.1} is the zoom in part of the year 1979. The red (dark blue) circles in \textbf{c}, blue circles in \textbf{d} and purple circles in \textbf{d} represent the one minus cumulative distribution of $\log$ waiting time of all the site records of air temperature differences for the positive, negative and absolute value quakes, respectively. The three plots follow an exponential, represented by the black lines.} \label{fig:atch_QA}
\end{figure}

Within the complex climate system, quakes are often associated with extreme events. Next, we study the evolution of the system by analysing the CDF of logarithmic waiting time between quakes. Here, we measure where the air temperature positive, negative differences ($\mathrm{CV}>0$, $\mathrm{CV}<0$) and the absolute value of all the changes ($\lvert \mathrm{CV} \rvert$) jump as positive value, negative value and absolute value quakes in Figs.~\ref{fig:atch_QA}a and b.

Figures.~\ref{fig:atch_QA}c-e show that the CDF of logarithmic waiting times follows  exponential forms for positive quakes, negative quakes and quakes, respectively. Our findings are highly consistent with the results in other complex systems \cite{anderson2004evolution}, revealing the air temperature differences of the weather system are ageing. That means it is harder to reach a new record, and the release of an intrinsic tension is decelerating even under global warming.

To further investigate the potential changing properties, we apply a moving time window of $3$ years to analyse whether the statistical characteristics vary with time. As the $\log(t_{k})$ is anticipated to follow a \textit{Poisson} distribution, the average number of events $\mathrm{d}\bar{n}$ per unit of time decays as~\cite{sibani2009record}:
\begin{equation}
    \frac{\mathrm{d}\bar{n}}{\mathrm{d}t} \propto \frac{1}{t}.
\end{equation}

\begin{figure}[!h]
    \centering
    \includegraphics[width=0.9\linewidth]{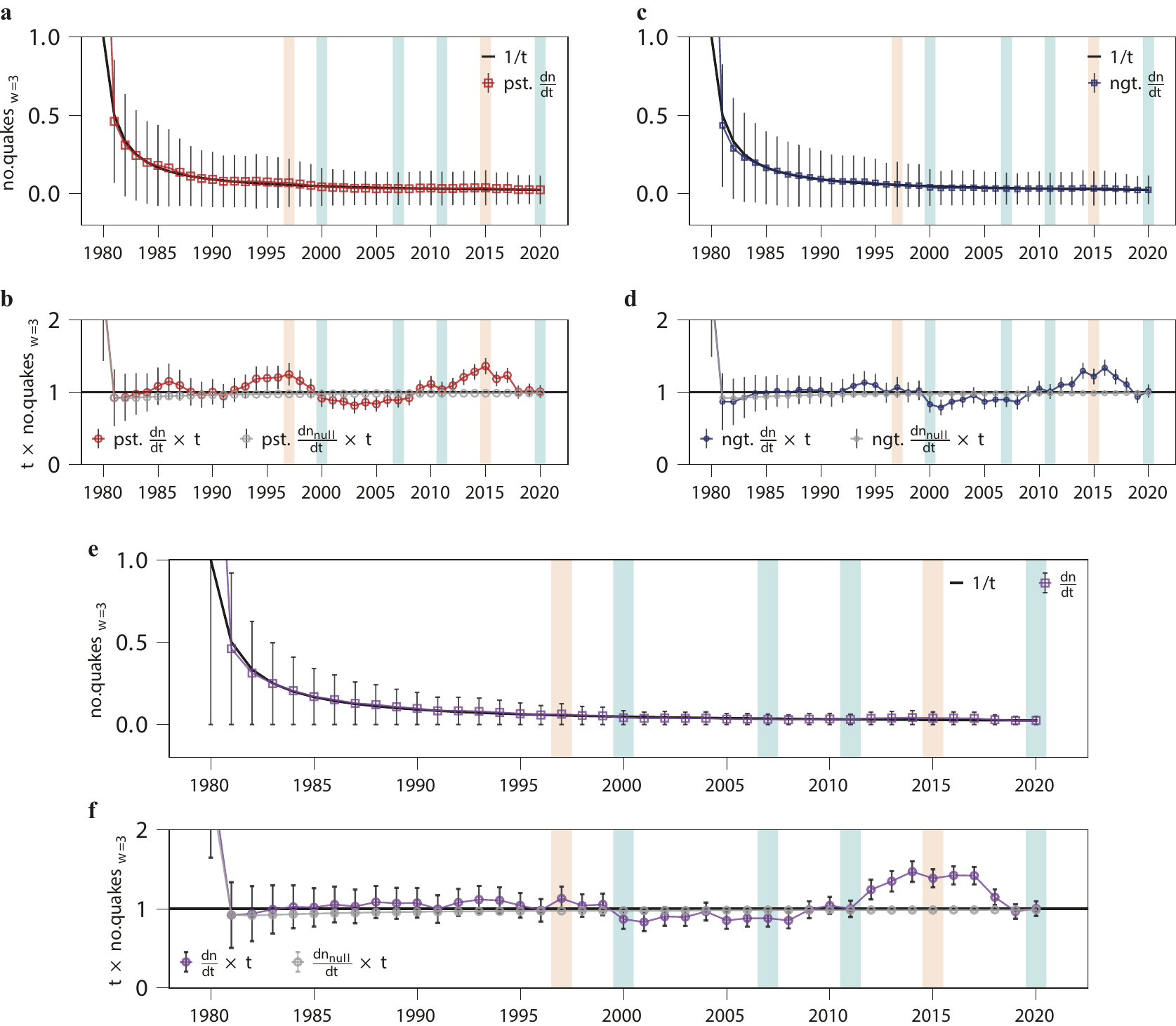}\caption{\textbf{The average number of quakes at a site in a unit time against the time in a linear-linear plot.} \textbf{a}, \textbf{c} and \textbf{e} plot the average number of quakes $\mathrm{\bar{n}}$ each site per year for $\mathrm{CV}>0$, $\mathrm{CV}<0$ and $\lvert\mathrm{CV}\rvert$ respectively. The quakes per year are averaged over a 3-year window time, and $\overline{n}$ is averaged over $N=10512$ sites. $1/t$ is plotted as a black line. \textbf{b}, \textbf{d} and \textbf{f} plot rescaled average number of the quakes of \textbf{a}, \textbf{c} and \textbf{e} against the time correspondingly. If $\mathrm{\bar{n}}$ decays as $1/t$, then the rescaled number should lie in a horizontal line. The error bars are indicated in black lines. This plot reveals clearly deviations away from a proposed scaling law from the year 2010 to 2019. The light orange and light green shadings indicate the \el~ and \la~ periods. We shuffle the dataset for $2000$ times to generate a null time series. The results are presented in grey with error bars smaller than the symbol size.}
    \label{fig:mw_quake}
\end{figure}
\clearpage

Figure~\ref{fig:mw_quake} demonstrates that the average number of quakes at each site in a three-year window time decays almost as one over time; however, in the rescaled plots $\frac{\mathrm{d}\bar{n}}{\mathrm{d}t} \times t = \text{constant}$, we uncover that there are some deviations
throughout the evolution, as shown in Figs.~\ref{fig:mw_quake}b, d and f. These deviations are usually related to the ENSO extreme events. We highlight the duration of the \el~ and \la~ events using the light orange and light green shading, respectively. We propose the underlying mechanism is that the ENSO events seem to impact the weather variability of both the Northern  and Southern Hemispheres~\cite{yang_impacts_2020}, leading to stronger baroclinic disturbances. This can cause stronger day-to-day time variability~\cite{machado_influence_2021} and make extreme weather events harder to predict. In particular, super and extreme \el~ events may become more frequent under climate change~\cite{wang_historical_2019,cai_increasing_2014}. Nevertheless, the overall trend is consistent with relaxational record dynamics. 




We further test the robustness of the PDF of daily temperature differences in the time window of $10$ years, as shown in Fig.~S9. We find that the distributions of magnitudes at the top $5\%$ are very similar, indicating that the magnitudes of the defined extreme events follow a steady trend. We perform the same analysis on the CMIP6 (Coupled Model Intercomparison Project Phase 6)~\cite{gmd-9-1937-2016}
model data as well. Detailed information about this dataset is summarised in the Data availability section. The robust results of the same scaling phenomena validate the universal scaling phenomenon is independent of the database, see Tab.~S4 in SI.

\section*{Discussion}
Widespread extreme weather gets a boost from anthropogenic climate change and leads to significant socioeconomic ramifications in natural and human systems. In this study, we highlight the extreme values of day-to-day temperature differences. This seldom studied climatic variable provides a measure of daily weather variability that is relevant in terms of impacts on human and environmental systems because rapid temperature variations - both positive and negative - can be extremely damaging. 

We apply statistical physics, extreme value theory and record statistics to reveal the universal scaling of magnitude and waiting time of extreme events as well as the record statistics of daily air temperature variability. We find that the distribution of the extreme magnitudes is characterised by the \textit{Gumbel} distribution, while the waiting time of these extreme events is governed by the generalised \textit{Gamma} distribution. This analysis is also confirmed by the extreme precipitation (see Figs.~S10 and S11) and CMIP6 model data (see Tab.~S4). In particular, in the CMIP6 global warming experiments, we find that our universal scaling is quite robust across different models under climate change. That is, the last $43$ (2058-2100) years data can be collapsed onto the distributions of the first $43$ (2015-2057) years.
Further scale analysis reveals that the universal scaling phenomenon is independent of the system sizes and regions. This can be explained by the renormalisation group theory of extreme values ~\cite{gyorgyi2010renormalization, calvo2012extreme, corral2015scaling}.

We further investigated the evolution of extreme events of day-to-day temperature variations by analysing the `quakes'. The cumulative distribution of logarithmic waiting time of the records in the considered data, 
obeys to a good approximation an exponential law, indicating the system's dynamics has a direction and does not deviate substantially from steady state conditions.
Note that this is in stark opposition with the case of daily temperatures, where records are being beaten more and more frequently as a result of climate change. 

While climate change seems not to impact the statistics of extreme day-to-day temperature variations, the most relevant deviation from the exponential law for the quakes is found in concurrence with 
ENSO events, when records are beaten more frequently than what steady state conditions would otherwise predict.  
After the end of the ENSO event, the statistical properties fall back to the stationary case. We discussed the underlying mechanism for this abnormality and associated it
 with the impacts on the weather variability of ENSO events. 
The analysis of quakes is a promising tool for investigating the future impact of climate change on various climatic variables, including different measures of weather variability. 
As mentioned above, in~\cite{Kotz2021b}, a clear signature of climate change was found on the intramonthly standard deviation of daily surface temperature. The same applies if one considers the temperature record itself: the ongoing climate change generates a strong bias in the time statistics of quakes. The result by~\cite{Kotz2021b} disagrees with ours, but, as mentioned above, we are using a rather different measure of daily temperature variability. 





Our findings extend our understanding of the nature of extreme weather events. They could help us design schemes for reducing risks associated with sudden temperature changes from one day to the next.  
Moreover, the universal scaling in extreme events might be applied to other complex systems, such as biological, ecological and financial systems~\cite{sornette2006critical}.


\section*{Methods}
\subsection*{Fitting methods}
\textit{Fitting with Maximum Likelihood Estimator}. The Maximum Likelihood Estimation (MLE) is a general and flexible method of estimation of the unknown parameters, and each set of the parameters attaches to different probability densities of the observed data. The probability of the empirical data as a function of parameters is called the likelihood function. Values of high likelihood parameters correspond to models that give high probability to the observed data.

\textit{Fitting with the least square errors}. This article uses the least square errors method to fit the lines for the collapsed waiting time data. Specifically, we use the Levenberg-Marquardt algorithm as a standard tool for nonlinear fitting. It combines two minimization methods, namely gradient descent and the Gauss-Newton method.

\subsection*{Goodness of fit test}
\textit{Kolmogorov–Smirnov goodness of fit test}.
We apply the Kolmogorov–Smirnov test (KS test) to the magnitudes, waiting times of the extreme events, and the quakes to assess the goodness of fit. The Kolmogorov–Smirnov statistic is defined as $D_n = \underset {x}{\sup} \lvert F_{n}(x) - F(x)\rvert$, where $F_{n}(x)$ and $F(x)$ are the empirical distribution function of the sample and the cumulative distribution function of the reference distribution respectively. This statistic quantifies the distance between those two distributions. By the Glivenko–Cantelli theorem, if the sample comes from distribution $F(x)$, then $D_n$ converges to 0 almost surely in the limit when the number of samples $n$ goes to infinity. When $\sqrt{n}D_n > K_{\alpha}$, where $\mathrm{Pr}(K \leq K_{\alpha}) = 1- \alpha$, the test can reject the null hypothesis (the sample comes from distribution $F(x)$) at the $1-\alpha$ confidence level.

\subsection*{Generalised Extreme Value Theory}
The probability density function of $\textrm{Generalized Extreme Value}, ~\mathrm{GEV}(\mu, \beta, \xi)$ distribution family is defined for the maximum (or minimum) values of random variables, mathematically:
\begin{flalign} \label{eq:1}
  p(z) &= \frac{1}{\beta}t(z)^{\xi + 1}e^{-t(z)}, \\ 
    t(z) &= \left\{\begin{matrix}
            (1+\xi (\frac{z-\mu}{\beta}))^{-\frac{1}{\xi}}& \xi \neq 0\\
            e^{-\frac{z-\mu}{\beta}}& \xi = 0 ,
            \end{matrix}\right. \nonumber
\end{flalign}
and the cumulative density function is:
\begin{equation}\label{eq:extreme}
  G(z) = \left\{\begin{matrix}
\exp  \left \{   -\left[ 1+ \xi (\frac{z-\mu}{\beta}) \right] ^{-1 / \xi}  \right \} & \xi \neq 0 \\ 
\exp  \left \{   - \exp \left[ (\frac{z-\mu}{\beta}) \right] \right \} & \xi = 0,
\end{matrix}\right.
\end{equation}
where $\xi$ is the shape parameter and it can be any real number, $\mu$ and $\beta$ are the local and scale parameters. When $\xi = 0 $, the distribution is the~\textit{Gumbel} distribution which is widely used in climate modelling~\cite{lucarini2016extremes}.

\subsection*{Peaks Over Threshold}
\textit{The Generalised Pareto Distribution} is a family distribution $H(y)$ defined for the excesses $y = \mathbf{X} - u$ of a sequence of independent random variables $X_{1},X_{2},X_{3}\cdots$ with common distribution family.

\textbf{Theorem}: $M_{n} = \mathrm{max} \left \{  X_1, \cdots, X_n\right \}, \mathrm{Pr}\left\{ M_n \leq z \right\} \approx G(z)$, where $G(z) = \exp  \left \{   -\left[ 1+ \xi (\frac{z-\mu}{\beta}) \right] ^{-1 / \xi}  \right \}$, for some $\mu,\beta > 0$ and $\xi \neq 0$ while $G(z) = \exp  \left \{   - \exp \left[ (\frac{z-\mu}{\beta}) \right] \right \}$ for some $\mu,\beta > 0$ and $\xi = 0$. $G(z)$ represents the GEV family. 
Then for large enough $u$, conditional on $X>u$, approximately,
\begin{equation}
    H(y) = 
\left\{\begin{matrix}
 1-\left( 1+\frac{\tilde{\xi} y}{\tilde{\sigma}} \right)^{1/\tilde{\xi}} & \tilde{\xi} \neq 0 ,\\ 
 1-  \exp \left( \frac{y}{\tilde{\sigma}} \right) & \tilde{\xi}  = 0 .
\end{matrix}\right.
\end{equation}
If convergence between the GEV and GP distributions is realised, $\xi = \tilde{\xi}$ and $\tilde{\sigma} = \beta + \xi(u - \mu)$. This theorem implies that the parameters of the generalized Pareto distribution of threshold excesses are uniquely determined by those of the associated GEV distribution $G$. As a result, once we estimate the parameters for the GP parameters, we can derive the corresponding GEV and vice versa. In particular, the $\xi$ for the two distributions are the same. The GP scale parameter depends on the chosen threshold, we usually look at the modified scale parameter which is defined as $\sigma = \tilde{\sigma}_{u} - \xi u$.

\subsection*{Record Statistics and Dynamics}
A record within an ordered series of numerical data obtained from observations and measurements is simply a value larger (or smaller) than all preceding values. The mathematical properties of records depend on the properties of the series from which they are generated. And, the record size fluctuations of, e.g. the energy can push the system to cross the `edge of stability, inducing irreversible configurational changes. The statistics of those changes closely follow the statistics of record fluctuations. 

Consider a sequence of independent random numbers drawn from the same distribution ar times $1$, $2$, $\cdots$,~$t$. When this time series is sufficiently long, the number of records found between the first and the $t$th draw has a Poisson distribution, with an average $\ln t$. That is:
\begin{equation}
    P_{n}(t) = \frac{(\ln t)^{n-1}}{(n-1)!}\frac{1}{t},
\end{equation}
where $P_{n}(t)$ is the probability of having $n$ records in $t$ successive trials, where $1 \leq n \leq t$. It is a Poisson distribution when replacing usual time argument $t$ with $\ln t$. Therefore, the differences between two Poisson processes, the ratios $\ln(t_{k}/t_{k-1})$ are independent random numbers with an exponential distributions, where $t_{k}$ is the time when the $k$th record appears. Let $\overline{n(t)}$ be the average number of vents in time t, we have $\overline{n(t)} = \ln t$. Furthermore, the average number of events per unit of time decays as: $\frac{\mathrm{d}\bar{n}}{\mathrm{d}t} \propto \frac{1}{t}$.     


\section*{Data availability}
The primary dataset is the Reanalysis Air temperature ERA5 (ECMWF Reanalysis 5th Generation) 1000hPa Daily (time:00:00) from 1979-2021. We deleted the data point of 29th Feb in any leap year~\cite{hersbach2020era5}. ECMWF is the European Centre for Medium-Range Weather Forecasts.
The CMIP6 (Coupled Model Intercomparison Project Phase 6) model data is provided by `The Program for Climate Model Diagnosis and Intercomparison'~\cite{gmd-9-1937-2016}.
The data spans from 2015 to 2100. More details of the models are described in SI. 

\section*{Acknowledgments}
The authors would like to thank Prof. J\"urgen Kurths, Dr. Xiaojie Chen and Teng Liu for many exchanges on extreme events and the access of dataset and appreciate Dr. Nanxin Wei's useful discussions on the scaling phenomena. The authors acknowledge the support of the National Natural Science Fundation of China (Grant No. 12135003) and Liyun Postdoctoral Program of Beijing Normal University. V.L. acknowledges the support received from the Horizon 2020 project TiPES (grant no. 820970) and from the EPSRC Grant EP/T018178/1.

{\section*{Author Contributions}
J.F., X.C. designed the research. Q.Y. performed the analysis and prepared the manuscript, Q.Y., J.M., J.F., V.L., H.J., K.C., X.C. generated research ideas and discussed results, and contributed to writing the manuscript.}

\section*{Additional information}
Supplementary Information is available in the online version of the paper.

\section*{Competing interests}
The authors declare no competing  interests.


\bibliographystyle{naturemag}
\bibliography{extreme_weather_variability-v2}

\begin{thebibliography}{10}
\expandafter\ifx\csname url\endcsname\relax
  \def\url#1{\texttt{#1}}\fi
\expandafter\ifx\csname urlprefix\endcsname\relax\def\urlprefix{URL }\fi
\providecommand{\bibinfo}[2]{#2}
\providecommand{\eprint}[2][]{\url{#2}}

\bibitem{odor2004universality}
\bibinfo{author}{{\'O}dor, G.}
\newblock \bibinfo{title}{Universality classes in nonequilibrium lattice
  systems}.
\newblock \emph{\bibinfo{journal}{Reviews of modern physics}}
  \textbf{\bibinfo{volume}{76}}, \bibinfo{pages}{663} (\bibinfo{year}{2004}).

\bibitem{fan2020universal}
\bibinfo{author}{Fan, J.} \emph{et~al.}
\newblock \bibinfo{title}{Universal gap scaling in percolation}.
\newblock \emph{\bibinfo{journal}{Nature Physics}}
  \textbf{\bibinfo{volume}{16}}, \bibinfo{pages}{455--461}
  (\bibinfo{year}{2020}).

\bibitem{munoz2018colloquium}
\bibinfo{author}{Munoz, M.~A.}
\newblock \bibinfo{title}{Colloquium: Criticality and dynamical scaling in
  living systems}.
\newblock \emph{\bibinfo{journal}{Reviews of Modern Physics}}
  \textbf{\bibinfo{volume}{90}}, \bibinfo{pages}{031001}
  (\bibinfo{year}{2018}).

\bibitem{di2018landau}
\bibinfo{author}{Di~Santo, S.}, \bibinfo{author}{Villegas, P.},
  \bibinfo{author}{Burioni, R.} \& \bibinfo{author}{Mu{\~n}oz, M.~A.}
\newblock \bibinfo{title}{Landau--ginzburg theory of cortex dynamics:
  Scale-free avalanches emerge at the edge of synchronization}.
\newblock \emph{\bibinfo{journal}{Proceedings of the National Academy of
  Sciences}} \textbf{\bibinfo{volume}{115}}, \bibinfo{pages}{E1356--E1365}
  (\bibinfo{year}{2018}).

\bibitem{sole2007scaling}
\bibinfo{author}{Sol{\'e}, R.}
\newblock \bibinfo{title}{Scaling laws in the drier}.
\newblock \emph{\bibinfo{journal}{Nature}} \textbf{\bibinfo{volume}{449}},
  \bibinfo{pages}{151--153} (\bibinfo{year}{2007}).

\bibitem{ross2021universal}
\bibinfo{author}{Ross, S. R.-J.} \emph{et~al.}
\newblock \bibinfo{title}{Universal scaling of robustness of ecosystem services
  to species loss}.
\newblock \emph{\bibinfo{journal}{Nature communications}}
  \textbf{\bibinfo{volume}{12}}, \bibinfo{pages}{1--7} (\bibinfo{year}{2021}).

\bibitem{song2010modelling}
\bibinfo{author}{Song, C.}, \bibinfo{author}{Koren, T.}, \bibinfo{author}{Wang,
  P.} \& \bibinfo{author}{Barab{\'a}si, A.-L.}
\newblock \bibinfo{title}{Modelling the scaling properties of human mobility}.
\newblock \emph{\bibinfo{journal}{Nature physics}}
  \textbf{\bibinfo{volume}{6}}, \bibinfo{pages}{818--823}
  (\bibinfo{year}{2010}).

\bibitem{bouchaud1990anomalous}
\bibinfo{author}{Bouchaud, J.-P.} \& \bibinfo{author}{Georges, A.}
\newblock \bibinfo{title}{Anomalous diffusion in disordered media: statistical
  mechanisms, models and physical applications}.
\newblock \emph{\bibinfo{journal}{Physics reports}}
  \textbf{\bibinfo{volume}{195}}, \bibinfo{pages}{127--293}
  (\bibinfo{year}{1990}).

\bibitem{mantegna1995scaling}
\bibinfo{author}{Mantegna, R.~N.} \& \bibinfo{author}{Stanley, H.~E.}
\newblock \bibinfo{title}{Scaling behaviour in the dynamics of an economic
  index}.
\newblock \emph{\bibinfo{journal}{Nature}} \textbf{\bibinfo{volume}{376}},
  \bibinfo{pages}{46--49} (\bibinfo{year}{1995}).

\bibitem{stanley2000scale}
\bibinfo{author}{Stanley, H.~E.} \emph{et~al.}
\newblock \bibinfo{title}{Scale invariance and universality: organizing
  principles in complex systems}.
\newblock \emph{\bibinfo{journal}{Physica A: Statistical Mechanics and its
  Applications}} \textbf{\bibinfo{volume}{281}}, \bibinfo{pages}{60--68}
  (\bibinfo{year}{2000}).

\bibitem{sornette2006critical}
\bibinfo{author}{Sornette, D.}
\newblock \emph{\bibinfo{title}{Critical phenomena in natural sciences: chaos,
  fractals, selforganization and disorder: concepts and tools}}
  (\bibinfo{publisher}{Springer Science \& Business Media},
  \bibinfo{year}{2006}).

\bibitem{west2017scale}
\bibinfo{author}{West, G.}
\newblock \emph{\bibinfo{title}{Scale: The Universal Laws of Growth,
  Innovation, Sustainability, and the Pace of Life in Organisms, Cities,
  Economies, and Companies}} (\bibinfo{publisher}{Penguin Press},
  \bibinfo{year}{2017}).

\bibitem{ashkenazy2003nonlinearity}
\bibinfo{author}{Ashkenazy, Y.}, \bibinfo{author}{Baker, D.~R.},
  \bibinfo{author}{Gildor, H.} \& \bibinfo{author}{Havlin, S.}
\newblock \bibinfo{title}{Nonlinearity and multifractality of climate change in
  the past 420,000 years}.
\newblock \emph{\bibinfo{journal}{Geophysical research letters}}
  \textbf{\bibinfo{volume}{30}} (\bibinfo{year}{2003}).

\bibitem{fan2021statistical}
\bibinfo{author}{Fan, J.} \emph{et~al.}
\newblock \bibinfo{title}{Statistical physics approaches to the complex earth
  system}.
\newblock \emph{\bibinfo{journal}{Physics reports}}
  \textbf{\bibinfo{volume}{896}}, \bibinfo{pages}{1--84}
  (\bibinfo{year}{2021}).

\bibitem{bak2002unified}
\bibinfo{author}{Bak, P.}, \bibinfo{author}{Christensen, K.},
  \bibinfo{author}{Danon, L.} \& \bibinfo{author}{Scanlon, T.}
\newblock \bibinfo{title}{Unified scaling law for earthquakes}.
\newblock \emph{\bibinfo{journal}{Physical Review Letters}}
  \textbf{\bibinfo{volume}{88}}, \bibinfo{pages}{178501}
  (\bibinfo{year}{2002}).

\bibitem{saichev2006universal}
\bibinfo{author}{Saichev, A.} \& \bibinfo{author}{Sornette, D.}
\newblock \bibinfo{title}{“universal” distribution of interearthquake times
  explained}.
\newblock \emph{\bibinfo{journal}{Physical Review Letters}}
  \textbf{\bibinfo{volume}{97}}, \bibinfo{pages}{078501}
  (\bibinfo{year}{2006}).

\bibitem{corral2004long}
\bibinfo{author}{Corral, {\'A}.}
\newblock \bibinfo{title}{Long-term clustering, scaling, and universality in
  the temporal occurrence of earthquakes}.
\newblock \emph{\bibinfo{journal}{Physical Review Letters}}
  \textbf{\bibinfo{volume}{92}}, \bibinfo{pages}{108501}
  (\bibinfo{year}{2004}).

\bibitem{livina_memory_2005}
\bibinfo{author}{Livina, V.~N.}, \bibinfo{author}{Havlin, S.} \&
  \bibinfo{author}{Bunde, A.}
\newblock \bibinfo{title}{Memory in the {Occurrence} of {Earthquakes}}.
\newblock \emph{\bibinfo{journal}{Physical Review Letters}}
  \textbf{\bibinfo{volume}{95}}, \bibinfo{pages}{208501}
  (\bibinfo{year}{2005}).

\bibitem{ghil2020physics}
\bibinfo{author}{Ghil, M.} \& \bibinfo{author}{Lucarini, V.}
\newblock \bibinfo{title}{The physics of climate variability and climate
  change}.
\newblock \emph{\bibinfo{journal}{Reviews of Modern Physics}}
  \textbf{\bibinfo{volume}{92}}, \bibinfo{pages}{035002}
  (\bibinfo{year}{2020}).

\bibitem{franzke2020structure}
\bibinfo{author}{Franzke, C.~L.} \emph{et~al.}
\newblock \bibinfo{title}{The structure of climate variability across scales}.
\newblock \emph{\bibinfo{journal}{Reviews of Geophysics}}
  \textbf{\bibinfo{volume}{58}}, \bibinfo{pages}{e2019RG000657}
  (\bibinfo{year}{2020}).

\bibitem{katz1992extreme}
\bibinfo{author}{Katz, R.~W.} \& \bibinfo{author}{Brown, B.~G.}
\newblock \bibinfo{title}{Extreme events in a changing climate: variability is
  more important than averages}.
\newblock \emph{\bibinfo{journal}{Climatic change}}
  \textbf{\bibinfo{volume}{21}}, \bibinfo{pages}{289--302}
  (\bibinfo{year}{1992}).

\bibitem{karl1995trends}
\bibinfo{author}{Karl, T.~R.}, \bibinfo{author}{Knight, R.~W.} \&
  \bibinfo{author}{Plummer, N.}
\newblock \bibinfo{title}{Trends in high-frequency climate variability in the
  twentieth century}.
\newblock \emph{\bibinfo{journal}{Nature}} \textbf{\bibinfo{volume}{377}},
  \bibinfo{pages}{217--220} (\bibinfo{year}{1995}).

\bibitem{Easterling2000}
\bibinfo{author}{Easterling, D.~R.} \emph{et~al.}
\newblock \bibinfo{title}{Climate extremes: Observations, modeling, and
  impacts}.
\newblock \emph{\bibinfo{journal}{Science}} \textbf{\bibinfo{volume}{289}},
  \bibinfo{pages}{2068--2074} (\bibinfo{year}{2000}).

\bibitem{Ipcc2012}
\bibinfo{author}{IPCC}.
\newblock \emph{\bibinfo{title}{Managing the Risks of Extreme Events and
  Disasters to Advance Climate Change Adaptation}}
  (\bibinfo{publisher}{Cambridge Univerity Presse}, \bibinfo{year}{2012}).

\bibitem{Allen2003}
\bibinfo{author}{Allen, M.}
\newblock \bibinfo{title}{Liability for climate change}.
\newblock \emph{\bibinfo{journal}{Nature}} \textbf{\bibinfo{volume}{421}},
  \bibinfo{pages}{891--892} (\bibinfo{year}{2003}).

\bibitem{trenberth2015attribution}
\bibinfo{author}{Trenberth, K.~E.}, \bibinfo{author}{Fasullo, J.~T.} \&
  \bibinfo{author}{Shepherd, T.~G.}
\newblock \bibinfo{title}{Attribution of climate extreme events}.
\newblock \emph{\bibinfo{journal}{Nature Climate Change}}
  \textbf{\bibinfo{volume}{5}}, \bibinfo{pages}{725--730}
  (\bibinfo{year}{2015}).

\bibitem{Otto2017}
\bibinfo{author}{Otto, F.~E.}
\newblock \bibinfo{title}{Attribution of weather and climate events}.
\newblock \emph{\bibinfo{journal}{Annual Review of Environment and Resources}}
  \textbf{\bibinfo{volume}{42}}, \bibinfo{pages}{627--646}
  (\bibinfo{year}{2017}).

\bibitem{WMO}
\bibinfo{author}{Organization, W.~M.}
\newblock \bibinfo{title}{Weather-related disasters increase over past 50
  years, causing more damage but fewer deaths}.
\newblock
  \bibinfo{howpublished}{https://public.wmo.int/en/media/press-release/weather-related-disasters-increase-over-past-50-years-causing-more-damage-fewer}
  (\bibinfo{year}{2021}).

\bibitem{Poumadere2005}
\bibinfo{author}{Poumad\'ere, M.}, \bibinfo{author}{Mays, C.},
  \bibinfo{author}{Le~Mer, S.} \& \bibinfo{author}{Blong, R.}
\newblock \bibinfo{title}{The 2003 heat wave in france: Dangerous climate
  change here and now}.
\newblock \emph{\bibinfo{journal}{Risk Analysis}}
  \textbf{\bibinfo{volume}{25}}, \bibinfo{pages}{1483--1494}
  (\bibinfo{year}{2005}).

\bibitem{shi2015impacts}
\bibinfo{author}{Shi, L.}, \bibinfo{author}{Kloog, I.},
  \bibinfo{author}{Zanobetti, A.}, \bibinfo{author}{Liu, P.} \&
  \bibinfo{author}{Schwartz, J.~D.}
\newblock \bibinfo{title}{Impacts of temperature and its variability on
  mortality in new england}.
\newblock \emph{\bibinfo{journal}{Nature climate change}}
  \textbf{\bibinfo{volume}{5}}, \bibinfo{pages}{988--991}
  (\bibinfo{year}{2015}).

\bibitem{wheeler2000temperature}
\bibinfo{author}{Wheeler, T.~R.}, \bibinfo{author}{Craufurd, P.~Q.},
  \bibinfo{author}{Ellis, R.~H.}, \bibinfo{author}{Porter, J.~R.} \&
  \bibinfo{author}{Prasad, P.~V.}
\newblock \bibinfo{title}{Temperature variability and the yield of annual
  crops}.
\newblock \emph{\bibinfo{journal}{Agriculture, Ecosystems \& Environment}}
  \textbf{\bibinfo{volume}{82}}, \bibinfo{pages}{159--167}
  (\bibinfo{year}{2000}).

\bibitem{kotz2021day}
\bibinfo{author}{Kotz, M.}, \bibinfo{author}{Wenz, L.},
  \bibinfo{author}{Stechemesser, A.}, \bibinfo{author}{Kalkuhl, M.} \&
  \bibinfo{author}{Levermann, A.}
\newblock \bibinfo{title}{Day-to-day temperature variability reduces economic
  growth}.
\newblock \emph{\bibinfo{journal}{Nature Climate Change}}
  \textbf{\bibinfo{volume}{11}}, \bibinfo{pages}{319--325}
  (\bibinfo{year}{2021}).

\bibitem{albeverio2006extreme}
\bibinfo{editor}{Albeverio, S.}, \bibinfo{editor}{Jentsch, V.} \&
  \bibinfo{editor}{Kantz, H.} (eds.) \emph{\bibinfo{title}{Extreme Events in
  Nature and Society}}.
\newblock The Frontiers Collection (\bibinfo{publisher}{Springer Berlin
  Heidelberg}, \bibinfo{year}{2006}).

\bibitem{lucarini2016extremes}
\bibinfo{author}{Lucarini, V.} \emph{et~al.}
\newblock \emph{\bibinfo{title}{Extremes and recurrence in dynamical systems}}
  (\bibinfo{publisher}{John Wiley \& Sons}, \bibinfo{year}{2016}).

\bibitem{Galfi2017}
\bibinfo{author}{G{\'a}lfi, V.~M.}, \bibinfo{author}{B{\'o}dai, T.} \&
  \bibinfo{author}{Lucarini, V.}
\newblock \bibinfo{title}{Convergence of extreme value statistics in a
  two-layer quasi-geostrophic atmospheric model}.
\newblock \emph{\bibinfo{journal}{Complexity}} \textbf{\bibinfo{volume}{2017}},
  \bibinfo{pages}{5340858} (\bibinfo{year}{2017}).

\bibitem{Faranda2017}
\bibinfo{author}{Faranda, D.}, \bibinfo{author}{Messori, G.} \&
  \bibinfo{author}{Yiou, P.}
\newblock \bibinfo{title}{Dynamical proxies of north atlantic predictability
  and extremes}.
\newblock \emph{\bibinfo{journal}{Scientific Reports}}
  \textbf{\bibinfo{volume}{7}}, \bibinfo{pages}{41278} (\bibinfo{year}{2017}).

\bibitem{Hochman2019}
\bibinfo{author}{Hochman, A.}, \bibinfo{author}{Alpert, P.},
  \bibinfo{author}{Harpaz, T.}, \bibinfo{author}{Saaroni, H.} \&
  \bibinfo{author}{Messori, G.}
\newblock \bibinfo{title}{A new dynamical systems perspective on atmospheric
  predictability: Eastern mediterranean weather regimes as a case study}.
\newblock \emph{\bibinfo{journal}{Science Advances}}
  \textbf{\bibinfo{volume}{5}}, \bibinfo{pages}{eaau0936}
  (\bibinfo{year}{2019}).

\bibitem{Galfi2021}
\bibinfo{author}{G{\'a}lfi, V.~M.}, \bibinfo{author}{Lucarini, V.},
  \bibinfo{author}{Ragone, F.} \& \bibinfo{author}{Wouters, J.}
\newblock \bibinfo{title}{Applications of large deviation theory in geophysical
  fluid dynamics and climate science}.
\newblock \emph{\bibinfo{journal}{La Rivista del Nuovo Cimento}}
  \textbf{\bibinfo{volume}{44}}, \bibinfo{pages}{291--363}
  (\bibinfo{year}{2021}).

\bibitem{helbing2013globally}
\bibinfo{author}{Helbing, D.}
\newblock \bibinfo{title}{Globally networked risks and how to respond}.
\newblock \emph{\bibinfo{journal}{Nature}} \textbf{\bibinfo{volume}{497}},
  \bibinfo{pages}{51--59} (\bibinfo{year}{2013}).

\bibitem{Leadbetter1983}
\bibinfo{author}{Leadbetter, M.~R.}, \bibinfo{author}{Lindgren, G.} \&
  \bibinfo{author}{Rootzen, H.}
\newblock \emph{\bibinfo{title}{Extremes and Related Properties of Random
  Sequences and Processes}}.
\newblock Springer Series in Statistics (\bibinfo{publisher}{Springer Verlag},
  \bibinfo{year}{1983}).

\bibitem{Katz2002}
\bibinfo{author}{Katz, R.~W.}, \bibinfo{author}{Parlange, M.~B.} \&
  \bibinfo{author}{Naveau, P.}
\newblock \bibinfo{title}{Statistics of extremes in hydrology}.
\newblock \emph{\bibinfo{journal}{Advances in Water Resources}}
  \textbf{\bibinfo{volume}{25}}, \bibinfo{pages}{1287--1304}
  (\bibinfo{year}{2002}).

\bibitem{Ghil2011}
\bibinfo{author}{Ghil, M.} \emph{et~al.}
\newblock \bibinfo{title}{Extreme events: dynamics, statistics and prediction}.
\newblock \emph{\bibinfo{journal}{Nonlinear Processes in Geophysics}}
  \textbf{\bibinfo{volume}{18}}, \bibinfo{pages}{295--350}
  (\bibinfo{year}{2011}).

\bibitem{coles2001introduction}
\bibinfo{author}{Coles, S.}, \bibinfo{author}{Bawa, J.},
  \bibinfo{author}{Trenner, L.} \& \bibinfo{author}{Dorazio, P.}
\newblock \emph{\bibinfo{title}{An introduction to statistical modeling of
  extreme values}}, vol. \bibinfo{volume}{208} (\bibinfo{publisher}{Springer},
  \bibinfo{year}{2001}).

\bibitem{Holton2004}
\bibinfo{author}{Holton, J.~R.}
\newblock \emph{\bibinfo{title}{{An introduction to dynamic meteorology}}}.
\newblock International Geophysics Series (\bibinfo{publisher}{Elsevier
  Academic Press,}, \bibinfo{address}{MA}, \bibinfo{year}{2004}),
  \bibinfo{edition}{4} edn.

\bibitem{Kotz2021b}
\bibinfo{author}{Kotz, M.}, \bibinfo{author}{Wenz, L.} \&
  \bibinfo{author}{Levermann, A.}
\newblock \bibinfo{title}{Footprint of greenhouse forcing in daily temperature
  variability}.
\newblock \emph{\bibinfo{journal}{Proceedings of the National Academy of
  Sciences}} \textbf{\bibinfo{volume}{118}}, \bibinfo{pages}{e2103294118}
  (\bibinfo{year}{2021}).

\bibitem{hersbach2020era5}
\bibinfo{author}{Hersbach, H.} \emph{et~al.}
\newblock \bibinfo{title}{The era5 global reanalysis}.
\newblock \emph{\bibinfo{journal}{Quarterly Journal of the Royal Meteorological
  Society}} \textbf{\bibinfo{volume}{146}}, \bibinfo{pages}{1999--2049}
  (\bibinfo{year}{2020}).

\bibitem{galfi2017convergence}
\bibinfo{author}{G{\'a}lfi, V.~M.}, \bibinfo{author}{B{\'o}dai, T.} \&
  \bibinfo{author}{Lucarini, V.}
\newblock \bibinfo{title}{Convergence of extreme value statistics in a
  two-layer quasi-geostrophic atmospheric model}.
\newblock \emph{\bibinfo{journal}{Complexity}} \textbf{\bibinfo{volume}{2017}}
  (\bibinfo{year}{2017}).

\bibitem{Zahid2017}
\bibinfo{author}{Zahid, M.}, \bibinfo{author}{Blender, R.},
  \bibinfo{author}{Lucarini, V.} \& \bibinfo{author}{Bramati, M.~C.}
\newblock \bibinfo{title}{Return levels of temperature extremes in southern
  pakistan}.
\newblock \emph{\bibinfo{journal}{Earth System Dynamics}}
  \textbf{\bibinfo{volume}{8}}, \bibinfo{pages}{1263--1278}
  (\bibinfo{year}{2017}).

\bibitem{boers2019complex}
\bibinfo{author}{Boers, N.} \emph{et~al.}
\newblock \bibinfo{title}{Complex networks reveal global pattern of
  extreme-rainfall teleconnections}.
\newblock \emph{\bibinfo{journal}{Nature}} \textbf{\bibinfo{volume}{566}},
  \bibinfo{pages}{373--377} (\bibinfo{year}{2019}).

\bibitem{pruessner2012self}
\bibinfo{author}{Pruessner, G.}
\newblock \emph{\bibinfo{title}{Self-organised criticality: theory, models and
  characterisation}} (\bibinfo{publisher}{Cambridge University Press},
  \bibinfo{year}{2012}).

\bibitem{Note1}
\bibinfo{note}{Note that in continental and arid areas, day-to-night excursions
  of more than 30 $^\circ C$ are far from atypical}.

\bibitem{meng_percolation_2017}
\bibinfo{author}{Meng, J.}, \bibinfo{author}{Fan, J.},
  \bibinfo{author}{Ashkenazy, Y.} \& \bibinfo{author}{Havlin, S.}
\newblock \bibinfo{title}{Percolation framework to describe {El} {Niño}
  conditions}.
\newblock \emph{\bibinfo{journal}{Chaos: An Interdisciplinary Journal of
  Nonlinear Science}} \textbf{\bibinfo{volume}{27}}, \bibinfo{pages}{035807}
  (\bibinfo{year}{2017}).

\bibitem{sibani1993slow}
\bibinfo{author}{Sibani, P.} \& \bibinfo{author}{Littlewood, P.~B.}
\newblock \bibinfo{title}{Slow dynamics from noise adaptation}.
\newblock \emph{\bibinfo{journal}{Physical review letters}}
  \textbf{\bibinfo{volume}{71}}, \bibinfo{pages}{1482} (\bibinfo{year}{1993}).

\bibitem{jensen2013stochastic}
\bibinfo{author}{Sibani, P.} \& \bibinfo{author}{Jensen, H.~J.}
\newblock \emph{\bibinfo{title}{Stochastic dynamics of complex systems: From
  glasses to evolution}}, vol.~\bibinfo{volume}{2} (\bibinfo{publisher}{World
  Scientific Publishing Company}, \bibinfo{year}{2013}).

\bibitem{anderson2004evolution}
\bibinfo{author}{Anderson, P.~E.}, \bibinfo{author}{Jensen, H.~J.},
  \bibinfo{author}{Oliveira, L.} \& \bibinfo{author}{Sibani, P.}
\newblock \bibinfo{title}{Evolution in complex systems}.
\newblock \emph{\bibinfo{journal}{Complexity}} \textbf{\bibinfo{volume}{10}},
  \bibinfo{pages}{49--56} (\bibinfo{year}{2004}).

\bibitem{sibani2009record}
\bibinfo{author}{Sibani, P.} \& \bibinfo{author}{Jensen, H.~J.}
\newblock \bibinfo{title}{Record statistics and dynamics}.
\newblock \emph{\bibinfo{journal}{Encyclopedia of Complexity and System
  Science}}  (\bibinfo{year}{2009}).

\bibitem{yang_impacts_2020}
\bibinfo{author}{Yang, M.}, \bibinfo{author}{Li, C.}, \bibinfo{author}{Tan,
  Y.}, \bibinfo{author}{Li, X.} \& \bibinfo{author}{Chen, X.}
\newblock \bibinfo{title}{Impacts of two types of {El}-{Niño} on the winter
  {North} {Pacific} storm track}.
\newblock \emph{\bibinfo{journal}{Environmental Research Letters}}
  \textbf{\bibinfo{volume}{15}}, \bibinfo{pages}{094062}
  (\bibinfo{year}{2020}).

\bibitem{machado_influence_2021}
\bibinfo{author}{Machado, J.~P.}, \bibinfo{author}{Justino, F.} \&
  \bibinfo{author}{Souza, C.~D.}
\newblock \bibinfo{title}{Influence of {El} {Niño}-{Southern} {Oscillation} on
  baroclinic instability and storm tracks in the {Southern} {Hemisphere}}.
\newblock \emph{\bibinfo{journal}{International Journal of Climatology}}
  \textbf{\bibinfo{volume}{41}}, \bibinfo{pages}{E93--E109}
  (\bibinfo{year}{2021}).

\bibitem{wang_historical_2019}
\bibinfo{author}{Wang, B.} \emph{et~al.}
\newblock \bibinfo{title}{Historical change of {El} {Niño} properties sheds
  light on future changes of extreme {El} {Niño}}.
\newblock \emph{\bibinfo{journal}{Proceedings of the National Academy of
  Sciences}} \textbf{\bibinfo{volume}{116}}, \bibinfo{pages}{22512--22517}
  (\bibinfo{year}{2019}).

\bibitem{cai_increasing_2014}
\bibinfo{author}{Cai, W.} \emph{et~al.}
\newblock \bibinfo{title}{Increasing frequency of extreme {El} {Niño} events
  due to greenhouse warming}.
\newblock \emph{\bibinfo{journal}{Nature Climate Change}}
  \textbf{\bibinfo{volume}{4}}, \bibinfo{pages}{111--116}
  (\bibinfo{year}{2014}).

\bibitem{gmd-9-1937-2016}
\bibinfo{author}{Eyring, V.} \emph{et~al.}
\newblock \bibinfo{title}{Overview of the coupled model intercomparison project
  phase 6 (cmip6) experimental design and organization}.
\newblock \emph{\bibinfo{journal}{Geoscientific Model Development}}
  \textbf{\bibinfo{volume}{9}}, \bibinfo{pages}{1937--1958}
  (\bibinfo{year}{2016}).

\bibitem{gyorgyi2010renormalization}
\bibinfo{author}{Gy{\"o}rgyi, G.}, \bibinfo{author}{Moloney, N.},
  \bibinfo{author}{Ozog{\'a}ny, K.}, \bibinfo{author}{R{\'a}cz, Z.} \&
  \bibinfo{author}{Droz, M.}
\newblock \bibinfo{title}{Renormalization-group theory for finite-size scaling
  in extreme statistics}.
\newblock \emph{\bibinfo{journal}{Physical Review E}}
  \textbf{\bibinfo{volume}{81}}, \bibinfo{pages}{041135}
  (\bibinfo{year}{2010}).

\bibitem{calvo2012extreme}
\bibinfo{author}{Calvo, I.}, \bibinfo{author}{Cuch{\'\i}, J.~C.},
  \bibinfo{author}{Esteve, J.~G.} \& \bibinfo{author}{Falceto, F.}
\newblock \bibinfo{title}{Extreme-value distributions and renormalization
  group}.
\newblock \emph{\bibinfo{journal}{Physical Review E}}
  \textbf{\bibinfo{volume}{86}}, \bibinfo{pages}{041109}
  (\bibinfo{year}{2012}).

\bibitem{corral2015scaling}
\bibinfo{author}{Corral, A.}
\newblock \bibinfo{title}{Scaling in the timing of extreme events}.
\newblock \emph{\bibinfo{journal}{Chaos, Solitons \& Fractals}}
  \textbf{\bibinfo{volume}{74}}, \bibinfo{pages}{99--112}
  (\bibinfo{year}{2015}).

\end{thebibliography}
\clearpage


\end{document}